\documentclass{article}

\usepackage{PRIMEarxiv}

\usepackage[utf8]{inputenc} 
\usepackage[T1]{fontenc}    

\usepackage{cite}
\usepackage{amsmath,amssymb,amsfonts}
\usepackage{algorithmic}
\usepackage{graphicx}
\usepackage{textcomp}
\usepackage{subcaption}

\usepackage{cite}
\usepackage{amsmath,amssymb,amsfonts}
\usepackage{algorithmic}
\usepackage{caption}
\usepackage{textcomp}
\usepackage{tabularx}
\usepackage{comment}
\usepackage{float}

\title{Emotion Recognition With Temporarily Localized 'Emotional Events' in Naturalistic Context
\thanks{\textit{\underline{Citation}}: 
\textbf{Authors. Title. Pages.... DOI:000000/11111.}} 
}

\author{
  Mohammad Asif \\
  Indian Institute of Information Technology Allahabad \\
  Prayagraj \\
  \texttt{pse2017001@iiita.ac.in} \\
   \And
   Sudhakar Mishra \\
Indian Institute of Information Technology Allahabad \\
  Prayagraj \\
  \texttt{rs163@iiita.ac.in} \\
  \And
  Majithia Tejas Vinodbhai \\
Indian Institute of Information Technology Allahabad \\
  Prayagraj \\
  \And
  Uma Shanker Tiwary \\
Indian Institute of Information Technology Allahabad \\
  Prayagraj \\
  \texttt{ust@iiita.ac.in} \\
}

\begin{document}
\maketitle

\begin{abstract}
Emotion recognition using EEG signals is an emerging area of research due to its broad applicability in BCI. Emotional feelings are hard to stimulate in the lab. Emotions don't last long, yet they need enough context to be perceived and felt. However, most EEG-related emotion databases either suffer from emotionally irrelevant details (due to prolonged duration stimulus) or have minimal context doubting the feeling of any emotion using the stimulus. We tried to reduce the impact of this trade-off by designing an experiment in which participants are free to report their emotional feelings simultaneously watching the emotional stimulus. We called these reported emotional feelings “Emotional Events” in our Dataset on Emotion with Naturalistic Stimuli (DENS). We used EEG signals to classify emotional events on different combinations of Valence(V) and Arousal(A) dimensions and compared the results with benchmark datasets of DEAP and SEED. STFT is used for feature extraction and used in the classification model consisting of CNN-LSTM hybrid layers. We achieved significantly higher accuracy with our data compared to DEEP and SEED data. We conclude that having precise information about emotional feelings improves the classification accuracy compared to long-duration EEG signals which might be contaminated by mind-wandering.
\end{abstract}


\section{Introduction}
\label{sec:introduction}
Emotion recognition has been a challenging task in artificial intelligence. Several methods are available for measuring the participants' emotions, such as behavioural changes and subjective experiences self-reported by the participants; peripheral and central nervous system measures\cite{mauss2009measures}, but brain activities are among the most robust dimensions of detecting human affect as it is difficult for the users to manipulate innate brain activity during the process. Accordingly, EEG is considered a suitable and convenient method to record electrical activities to measure brain activities as it is a non-invasive method, i.e. there are no scalpel incisions. 

Many studies have already been conducted to measure human affect with the help of EEG and other peripheral responses\cite{koelstra2011deap, zheng2015investigating, katsigiannis2017dreamer, miranda2018amigos}. In the previous studies, the focus of the study was to develop a database that is labelled and suitable for emotion detection by intelligent systems and has contributed to affective computing. There is a typical method in these studies to elicit emotion in the participants by presenting them with video clips as stimuli. In the process of emotion recognition and other classification tasks, all the EEG data for that stimulus has to be considered for the classification model as there is no information about the precise temporal location at which a participant may experience the emotion. Models must consider all the data presented for that label, which is unnecessarily computationally expensive and decreases the system's efficiency by feeding not-so-essential data in the input. 

In our approach, we have presented a novel method to overcome this issue by providing precise information about the emotion elicitation, self-reported by the participants. We call it an ‘Emotional Event’. In this method, an additional task is given to the participants to mention precise temporal information by clicking on their computer screens while watching the emotional clips if they feel some emotion.
To the best of our knowledge, there are also no EEG affective datasets available for the Indian subcontinent population. Hence we tried to fulfil this research gap also in our work. We have considered DEAP dataset\cite{koelstra2011deap} and SEED dataset\cite{zheng2015investigating} to be benchmark. We tried to follow the format similar to the benchmark datasets and compared our dataset’s results with these datasets based on statistical significance.

EEG measures the electrical signals from the scalp with temporal details. To detect emotions from brain activity using EEG, participants wear an EEG cap that captures the brain's electrical activity from the scalp through the electrodes. These electrodes are fixed with the help of an elastic cap at a specified distance from each other. Different EEG devices vary with the number of electrodes; that is called the number of channels of the EEG. Thirty-two or fewer EEG channels are especially notable in affective computing research\cite{rahman2021recognition}. There are also a few studies with up to 64 electrodes. In this work, we used 128 channels EEG device to detect emotions. This EEG cap follows the International 10-10 system’s standards\cite{luu2005determination}.

Emotions are complex and challenging to understand as many theories exist about emotions, and there is a lack of a single consensus\cite{izard2013human}. Study of emotions has been an emerging topic that combines multidiscipline fields of psychology, neuroscience, computer science and medicines, not limited to only these fields. There are different aspects involved in determining emotions, such as behavioural, psychological and physiological aspects, cognitive appraisals, expression through facial or vocal responses, and the subjective experience, to be last but not the least. This study focuses on physiological aspects of emotion, which are considered into account by the brain signals captured through EEG while watching emotional video clips. Further, this study tries to collect a comprehensive list of subjective experiences through a self-assessment rating at the end of each clip.

Many approaches could be used to assess the participants' emotional states. Earlier, some basic emotions were used that are universally recognised for study purposes\cite{izard2007basic}. Later, more theories came about some complex emotions that are a combination of basic emotions\cite{plutchik1980general}. Multi-dimensional theories of emotions are among the most notable standards for assessing core affect\cite{russell2003core, verma2014multimodal}. According to these theories, emotions are considered a multi-dimensional array; one dimension for valence (experiencing positive or negative) and the other for arousal (experiencing the intensity) or dominance (controlling or feeling controlled). There are also a few more dimensions that make the spectrum broader, e.g., relevance (), familiarity () and liking (). Asking participants to report these experiences on a continuous scale is common in similar studies. Some theories deal with the physiological responses of feeling emotions, e.g., body temperature and heartbeat change\cite{dewey1894theory, cannon1927james}. As viewing different theories, it is obvious to consider that emotion is not a single phenomenon; instead, it is a combination of physiological responses and other information. Evidence is available that many brain regions are involved during emotion perception\cite{dalgleish2004emotional}. We have collected ECG data of the participants along with EEG to consider these parameters. 

\textbf{Emotional Event:} Emotion is a complex phenomenon which is embedded within a context. Moreover, emotion is transient in nature and is not available throughout the stimulus duration. In fact, more than one aspect could be embedded within the stimulus context, and different participants can feel emotion at different points of time considering various aspects. However, most of the datasets recorded to date \cite{koelstra2011deap} \cite{zheng2015investigating} ignore the transient nature of emotions and provide a single emotional category for the whole stimulus duration. Although the stimulus has emotional information, it has some non-emotional aspects too, which could lead to mind wandering activity. Although there are some attempts to get continuous subjective feedback on emotional experience and neural activity, the experimental method involved multiple watching of the stimulus and retrospective collection of emotional experience \cite{zhang2016neural, young2017dynamic, raz2016functional, sachs2020dynamic, lettieri2022default}. The retrospective collection depends on autobiographical memory and can raise biases across subjects depending on their capability to recall\cite{andric2016repeated}. Also, repetitive viewing effects can bias the ratings and underlying neural effects\cite{andric2016repeated}. Hence, an experimental paradigm is needed to record the participants' feedback dynamically, with minimal distraction during emotion processing and minimizing the memory recall biases. To the best of our knowledge, in this work, we are introducing a novel paradigm in which the time-stamp of emotional feelings can be marked online that can be further utilized to get the subjective feedback of emotional feelings and analyze brain signals temporally localized to the feeling of an emotion. We will refer to these time-stamped emotional feelings as “emotional events”.

Emotion recognition through EEG data follows a similar pattern as follows in various EEG signal analyses. First, the data is acquainted, and some preprocessing is applied to the signal. These preprocessing steps involve removing artefacts such as ocular activity, muscle activity, and powerline interference. Also, downsampling of the signal and band-pass filtering are used to make data more useful. Various dimensionality reduction techniques such as ICA and PCA are also used to prune the data to make it feature-rich. After pre-processing, features are extracted from the signal to feed into the model for the classification task. There are different kinds of features available to extract that mainly include time-domain (e.g., event-related potential (ERP), high-order crossing (HOC), etc.), frequency-domain (e.g., power spectral density (PSD), etc.); and time-frequency domain (e.g., STFT, wavelet analysis, etc.)

EEG records multi-frequency non-stationary brain signals from various electrodes. Analyzing these signals is challenging because of the complex and irregular nature of EEG signals. The time-frequency domain analysis benefits both the time and frequency domains, e.g., better spatial and temporal information from EEG signals. One basic time-frequency domain feature extraction method is Short-Time Fourier Transform (STFT). STFT is a time-ordered sequence of spectral estimates and is one of the powerful and general-purpose signal processing techniques. It has been used in the field of spectral analysis of a signal. The STFT is used to compute spectrograms which are used extensively for signal processing. Spectrograms are visual representations of the spectrum of frequencies of a signal with varying times\cite{allen1982applications}.

CNN is the most frequently used architecture for EEG analysis and classification tasks, and DBN and RNN follow it\cite{roy2019deep}. Hence we have also used a hybrid combination of the CNN-LSTM model to train our data as it should be suitable to compare our results with benchmark datasets.
Using artificial intelligence for affective computing provides better learning capabilities to intelligent systems. With the advancement of computing power and the development of effective and advanced neural network research, the trend of using various machine learning and deep learning techniques has grown within the last few years\cite{ketkar2017deep}. This article employs the widely used state-of-the-art deep learning methods used to detect emotions from EEG signals.

\section{Methods}

\begin{figure*}
\centering
 \includegraphics[width=\linewidth]{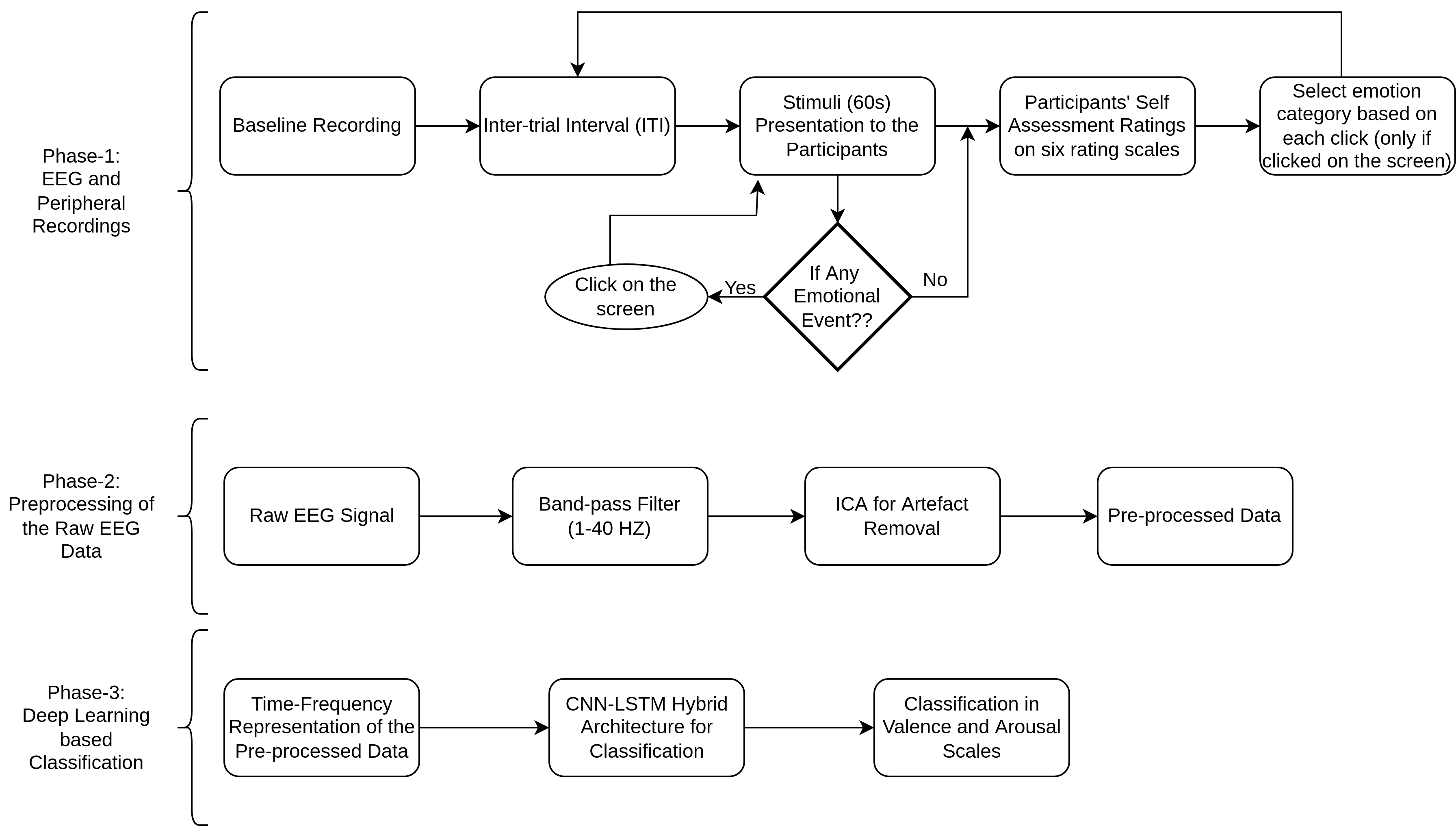}
  \caption{Complete Flowgram of the Experiment}
  \label{fig:FlowDiagram}
\end{figure*}

\subsection{Experimental Details}
The complete flow diagram of the experiment is given in Fig. \ref{fig:FlowDiagram}: Phase-1. We call our dataset 'Dataset on Emotion with Naturalistic Stimuli' (abbreviated as 'DENS'). 

\subsubsection{Stimuli}
The selection of stimuli to induce participants' emotions also plays a vital role in emotion recognition. A careful selection of stimuli is critical, and for that, technical validation of the video clips is crucial to assess if the intended emotional experience is elicited by the stimuli. In our previous work we have validated set of multimedia stimuli \cite{}. We selected 15 stimuli to perform our EEG experiment from the multimedia dataset. The table-\ref{tab:stimuli} shows list of stimuli with the target emotions assigned during the stimuli validation by \cite{}. 

\begin{table}[h]
\centering
\begin{tabular}{|p{5.5cm}|c|}
\textbf{Stimulus Name}  & \textbf{Target Emotion} \\ \hline
ashayen (199) & Adventorous\\
horror (214) & Afraid     \\
Anacondas The Hunt for the Blood Orchid clip (10) & Alarmed    \\
Lage Raho Munnabhai Only the Funny Scenes & Amused (98) \\
anger legend of baghat singh (198) & Angry      \\
Divergent Kiss scene clip& Aroused (54)   \\
Butterfly Nets (40) & Calm       \\
Best Horror Kills Ghost Ship Opening scene (26) & Disgust    \\
Jai Ho (92) & Enthusiastic       \\
SaddaHaq (152) & Excited    \\
MASOOM (109) & Happy      \\
cheerfulRang (201) & Joyous     \\
Crash Saddest scene (51) & Melancholic\\
hate lbs (210) & Miserable  \\
Madari movie of best scene (113) & Sad\\
Final Race of Milkha Singh Career (67) & Triumphant  \\ \hline
\end{tabular}%
\caption{Selected stimuli for EEG study from the stimuli dataset \cite{mishra2021affective}). The~time duration of each stimulus is 60 s. Bracket contains stimulus Id (available in open science framework repository).}
    \label{tab:stimuli}
\end{table}

\subsubsection{EEG Recording}
We recorded the EEG activity of forty participants ($23.3\pm 1.25$, F=3) while they were watching emotional film stimuli. The complete experimental procedure is available elsewhere \cite{mishra2022cardiac}. We are describing here the unique information which is relevant to this study. Each participant saw nine emotional stimuli, which were randomly extracted from the set of 16 stimuli. While watching the emotional film stimuli participants were instructed to perform a mouse click the moment they felt any emotion. At the end of each video stimuli, participants are provided six self-assessment scales, including valence, arousal, dominance, liking, familiarity, and relevance. Following these responses, for each click, participants were provided with a list of emotions pooled into four quadrants of V-A space (HVHA, LVHA, LVLA, HVLA) in the drop-down menu. Participants were also given a choice to enter the emotional category which suits their emotional experience but is unavailable in the provided emotion list. After that, an inter-trial-interval next trial started.

\subsubsection{Ratings}
\begin{figure*}
    \centering
    \includegraphics[width=\linewidth, height=0.25\linewidth]{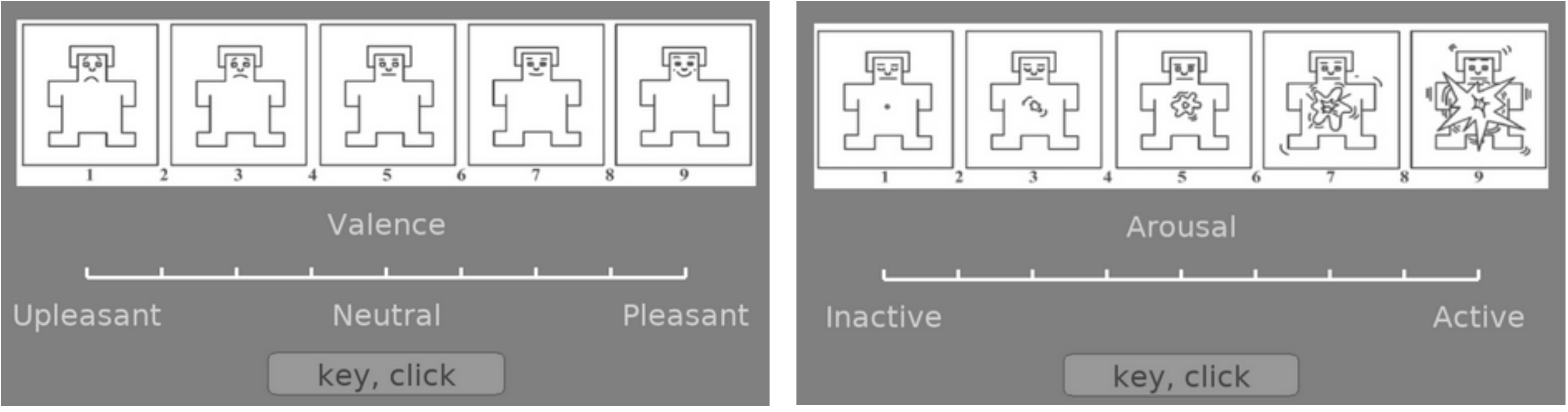}
    \caption{Valence and arousal rating scales used in the experiment.}
    \label{fig:ratingScales}
\end{figure*}

Subjective ratings are one of the well-known methods to evaluate the personal emotional experience of the participants. Emotional pictures/videos or audio clips are presented to the participants, and they are asked to rate these clips on different scales based on their personal experiences. These scales include Valence, Arousal, Dominance, Liking, Familiarity and Relevance. Although, in this analysis we considered only valence and arousal scales (represented in figure-\ref{fig:ratingScales}).

\subsection{Preprocessing and Artifact Removal of the EEG Data}
The procedure followed to perform the pre-processing is described elsewhere \cite{mishra2022cardiac}. The critical step which should be described here includes filtering and artifact removal. We had 128-channel EEG raw data with a sampling rate of 250 Hz. The raw signal is filtered using a Butterworth fifth-order bandpass filter with the passband 1-40 Hz. Independent component analysis (ICA) is used to remove artifacts, including heart rate, muscle movement, and eye blink-related artifacts.

\subsection{Other Datasets Used}
We have used DEAP dataset (a dataset for emotion analysis using eeg, physiological and video signals) and SEED dataset (A dataset collection for various purposes using EEG signals) for comparing the reults with our dataset (DENS).

The DEAP dataset consisted of 40 videos/trials, and for each trial there are 40 channels of EEG including peripheral signals are available and data is given foe each channel. We have used only 32 channels (i.e., discarded peripheral signals) for the experiment as we only want to use data from the brain only. This data was already pre-processed as 128 Hz downsampled, bandpass frequency of 4-45 Hz and EOG removed. For each trial, there are 4 labels available- Valence (V), Arousal (A), Dominance and Linking. We have used only V-A space for the experiment purpose.

The SEED dataset was recorded for 15 participants and emotions were presented to the participants into three categories- positive, negative and neutral emotions. We have used only V-space in DENS dataset to match the number of classes for both the datsets. The data was recorded using 62 channels.

\subsection{Feature Extraction}
EEG Signals are non-stationary, meaning the signal’s statistical characteristics change over time. If these signals are transformed to the frequency domain using Fourier Transform, it provides the frequency information, which is averaged over the entire EEG signal. So, information on different frequency events is not analyzed properly. If a signal is cut into minor segments such that it could be considered as stationary and focus on signal properties at a particular section which is called a windowing section and apply Fourier transform on it, it is called as Short-Time Fourier Transform (STFT).  It will move to the entire signal length and apply Fourier transform to find the spectral content of that section and display the coefficient as a function of both time and frequency. It provides insight into the nature of the time-varying spectral characteristics of the signal. Before STFT, let’s look at the discrete Fourier transform. Consider x: [0: L-1] = $\left \{ 0, 1, …, L-1 \right \}$ $\to$ R be a discrete-time signal where L is the signal length which is acquired by equidistance sample points with respect to the fixed sampling frequency. Mathematically DFT equation is,
\begin{equation} 
\hspace*{\fill}\widehat{x}(k)=\sum_{n=0}^{N-1} x(n).e^{-i2\pi nk/N}
\hspace*{\fill} \\
\end{equation}

Where k $\epsilon$ [0, K] and K is the frequency index with respect to Nyquist frequency. N is the duration of the section. The equation returns the complex Fourier coefficients for the k$^{th}$. These coefficients provide two parameters: phase and magnitude. For STFT, consider the additional parameter hop size (H), which is the step size of the window to be shifted. $\omega$ be a sampling window function which is $\omega$: [0, N-1] $\to$ R. STFT can be defined as,
\begin{equation} 
\hspace*{\fill}S(m,k):= \sum_{n=0}^{N-1} x(n+mH)\omega(n)e^{(-i2\pi kn/N)}\hspace*{\fill} \\
\end{equation}

Where m $\epsilon$ [0,M] and M is the maximum frame index mathematically M = $\left \lfloor \frac{L-N}{H} \right \rfloor$. The Short-Time Fourier Transform is not only a function of k but also m which is a proxy time representation. Here, the function returns Fourier coefficients for the k$^{th}$ proxy frequency at the m$^{th}$ temporal bin.\newline

Spectrograms are nothing but squared magnitude of STFT of the signal.
\begin{equation} 
\hspace*{\fill}\chi(m,k) := \left | S(m,k) \right |^2\hspace*{\fill} \\
\end{equation}
It is a 2D image where the horizontal axis represents time, and the vertical axis represents frequency bins. Number of frequency bins is (framesize / 2) + 1. No of time frames are ((size of signal – framesize) / hopsize) + 1. $\chi(m,k)$ represent intensity or color at (m, k).

\subsection{Input Preprocessing To Feed Data into The Classifier}
It is essential to convert the data into meaningful format that can be fed into our classifier model. As all the three datasets are available in different formats, we have provided information of the input preprocessing for each datasets as following:

\subsubsection{For DEAP Data:}

Each subject in the DEAP dataset is given by a tensor that is in the form of X $\in$ $\mathbb{R}^{40\times40\times8064}$, representing 40 videos, 40 EEG-channels (including peripheral channels), and 8064 EEG data samples for each channel. For labels, DEAP data provided a matrix in the form of X $\in$ $\mathbb{R}^{40\times4}$; i.e, for each subject, there 40 videos and 4 scales. From the DEAP dataset, the first 15 subjects are picked. For the label we have used Valence-Arousal space and divided it into four classes- HVHA, HVLA, LVLA, LVHA (abbreviations- H:High, L:Low, V:Valence, A:Arousal). We have converted 15 subjects data tensor into a matrix of X $\in$ $\mathbb{R}^{19200\times8064}$ $(i.e., 15 subjects \times 40 videos \times 32 channels, 8064 samples)$. Moreover, this data was processed for feature extraction using STFT with a window size of 0.5s and an overlap of 0.25s of data samples. Using STFT, we have converted every 8064 size of EEG data samples into a spectrogram image size of (33,251), as mentioned in the feature extraction section. Then, a hybrid CNN-LSTM classifier was implemented for multi-class classification with an input tensor of X $\in$ $\mathbb{R}^{33\times251\times3}$.

\subsubsection{For SEED Data:}

SEED dataset contains 45 .mat files for 15 subjects for each subject with 3 trials. The label file contains 3 emotional labels -1 for negative, 0 for neutral, and 1 for positive on the valence scale. After renaming, the labels become 0 for neutral, 1 for positive, and 2 for negative. For classification, we have considered 15 .mat files, one trial per subject. Due to the different sizes of data length in each channel, the first 16000 sample for each data which is the first 80s of data, is considered for further processing. EEG cap includes 62 channels according to the 10-20 international system. So, 15 subjects, 15 trials, 62-channels, and 16000 EEG data are converted into a tensor of X $\in$ $\mathbb{R}^{13950\times16000}$ $(i.e., 15 subjects \times 15 trials \times 62 channels, 16000 samples)$for feature extraction. As mentioned in the DEAP dataset experiment, using STFT with window size 0.5s and overlap of 0.25s, each 16000 EEG data is converted into a spectrogram with the shape of (51,319). Then, a hybrid CNN-LSTM classifier was implemented for multi-class classification with input tensor shape X $\in$ $\mathbb{R}^{51\times319\times3}$.

\subsubsection{For DENS Data:}

For the DENS dataset, we have 465 .mat files which contain emotional events. All the 465 files are picked for the experiment. Each .mat file is a matrix of X $\in$ $\mathbb{R}^{128\times1751}$, where 128 is the number of EEG channels and 1751 is the sample data for each channel. Then we have converted the data tensor of X $\in$ $\mathbb{R}^{465\times128\times1751}$ into the form of $\mathbb{R}^{59520\times1751}$ $(i.e., 465 emotional events \times 128 channels, 1751 samples)$ for feature extraction with window size 0.5s and overlap is 0.25s. After feature extraction, we have 59520 spectrograms, and each spectrogram is in the shape of (63, 26).
Now to compare with the DEAP dataset DENS dataset with 4 label classification is performed with a hybrid CNN-LSTM classifier. For the label, we have used the same V-A space (HVHA, HVLA, LVLA, LVHA) with input tensor of X $\in$ $\mathbb{R}^{63\times26\times3}$. For SEED and DENS datasets comparison, we have used 3-label classification. For the DENS dataset on the valence scale rating below 4.5 is marked as negative (0 labelled), and ratings above 5.5 is marked as positive (2 labelled). For neutral labels, in the DENS dataset, we have non-emotional files; we have marked neutral (1 labelled) for those files' data. Then with classifier, input tensor of X $\in$ $\mathbb{R}^{63\times26\times3}$ is used for classification.

\subsection{Model Architecture For the Classification Task}

Convolutional Neural Networks (CNN) and Long Short-Term Memory (LSTM) are one of the most widely used deep learning techniques. CNNs are used to extract meaningful patterns and features from the data. The key element in CNN is the convolution operation using kernels that automatically learn the local patterns from data. These local features are then combined into more complex features when multiple CNN layers are stacked. Filters (i.e, weights trained) in this process are also known as feature detector matrices. Input data will be convoluted with a filter map by sliding the kernel window. At the same time, LSTM networks can capture the sequential pattern as LDTMs are best suited for time-series data. LSTMs are designed to work for temporal correlations. Therefore, to exploit the benefits of both CNN and LSTM, a hybrid CNN-LSTM architecture is used for the classification of emotions.
The hybrid CNN-LSTM model utilizes the ability of convolutional layers for feature extraction from data, and LSTM layers are for long-term and short-term dependencies. The same model is used to compare all three datasets. The model classifier and it's details are shown in Fig. \ref{fig:model_arch}:

\begin{figure*}
\centering
 \includegraphics[width=\linewidth]{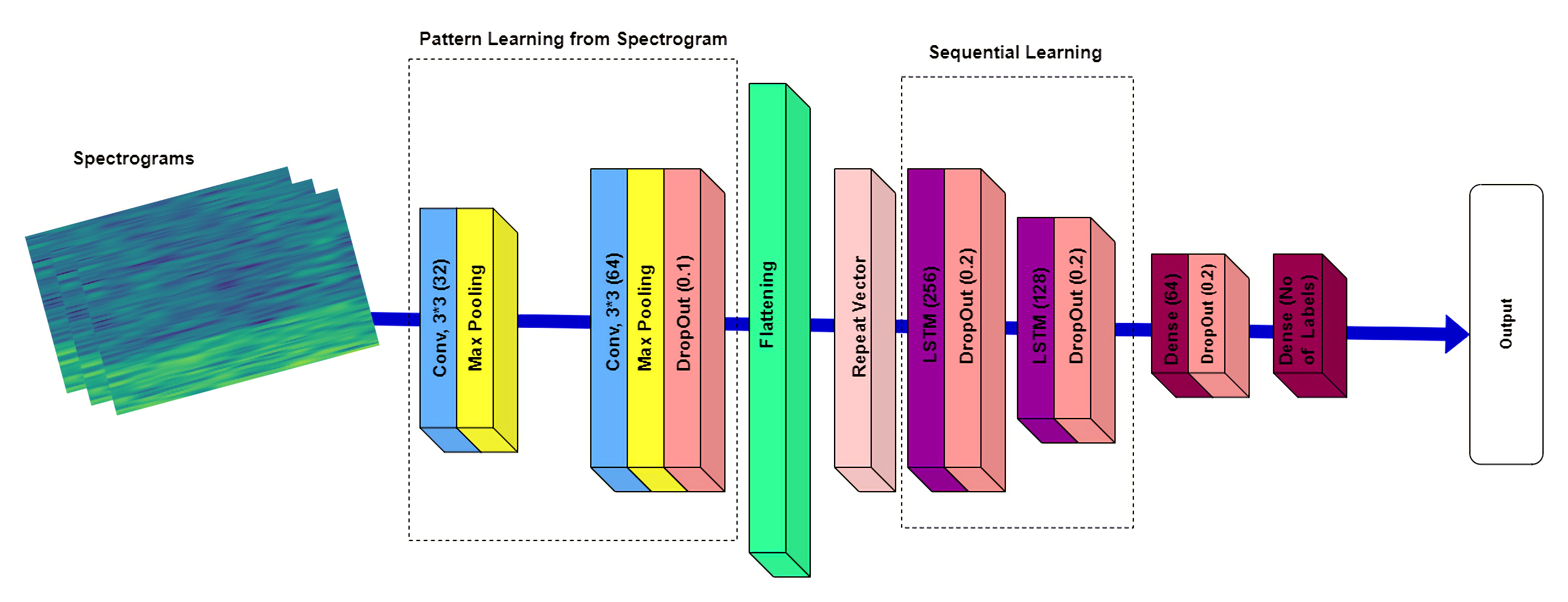}
  \caption{Model Architecture: It is consisted of two 2D-convolution layers with $3\times3$ kernels and 32 filters and 64 filters respectively, followed by a max pooling layer followed by a dropout layer and flattening layer. A repeat vector layer of size 4 is used before sending the data to the LSTM layers. Two LSTM layers are used of sizes 256 units and 128 units respectively, each followed by a dropout layer. At the end two dense layers are used of sizes 64 (followed by a dropout layer) and 4 or 3 (equals the number of the output classes).}
  \label{fig:model_arch}
\end{figure*}

CNN are often placed in the initial layers as it helps in local pattern learning from spectrogram or in general input data. The Pattern learning block consists of two 2D-convolutional blocks, each with a kernel size of $(3 \times 3)$. The feature map, which is the output of convolutional layers, keeps track of the location of the features in the input. A max-Pooling layer is added in between two consecutive convolutional layers. A pooling layer is added after the convolutional layer to reduce the feature-map dimension; hence it reduces the computational cost, and the activation function is applied to enhance the capability of the model. Rectified Linear Unit (ReLU) activation function which has been widely used to resilient vanishing gradient problem. In between, the dropout layer is used in some places to avoid the overfitting problem. 
The flattening layer transforms these feature maps into one-dimensional vectors. Repeat vector gives extra dimension for LSTM layer. The sequential learning block consists of 2 LSTM layers which capture the long-term temporal dependencies from the feature map extracted by CNN layers. 1$^{st}$ LSTM layer consists of 256 cells with a return sequence is set to ‘True’ while 2$^{nd}$ LSTM consists of 128 cells and as it is the last LSTM layer return sequence is ‘False’. Between LSTM layers, dropout layers with rate = 0.2 are added to avoid overfitting issues. Finally, two fully-connected layers where 1$^{st}$ layer with 64 neurons and 2$^{nd}$ layer with the number of classes as neurons are added for further processing. As we have the multi-class classification, the SoftMax activation function is used in the output layer as it outputs a vector representing the probability distributions of a list of potential classes.
 
The parameter setting for developed deep learning model is mentioned in the Table \ref{tab:Parameters}. 

\begin{table}
\centering
\begin{tabular}{|l|l|}
\hline
Parameter     & Setting                                   \\ \hline
Optimizer     & Adam                                      \\ \hline
Loss function & Categorical Cross-entropy                 \\ \hline
Learning rate & 0.001                                     \\ \hline
Adjustment    & \begin{tabular}[c]{@{}l@{}}Early Stopping criteria: monitor - 'val\_loss', patience = 30\\ Model Checkpoint: monitor - 'val\_accuracy'\end{tabular} \\ \hline
Batch size    & 256                                  \\ \hline
Epochs        & 100                                 \\ \hline
\end{tabular}
\caption{Parameter Settings for the Model}
    \label{tab:Parameters}
\end{table}

\section{Results}
\begin{figure}
\centering
\begin{subfigure}{0.49\textwidth}
  \includegraphics[width=\textwidth]{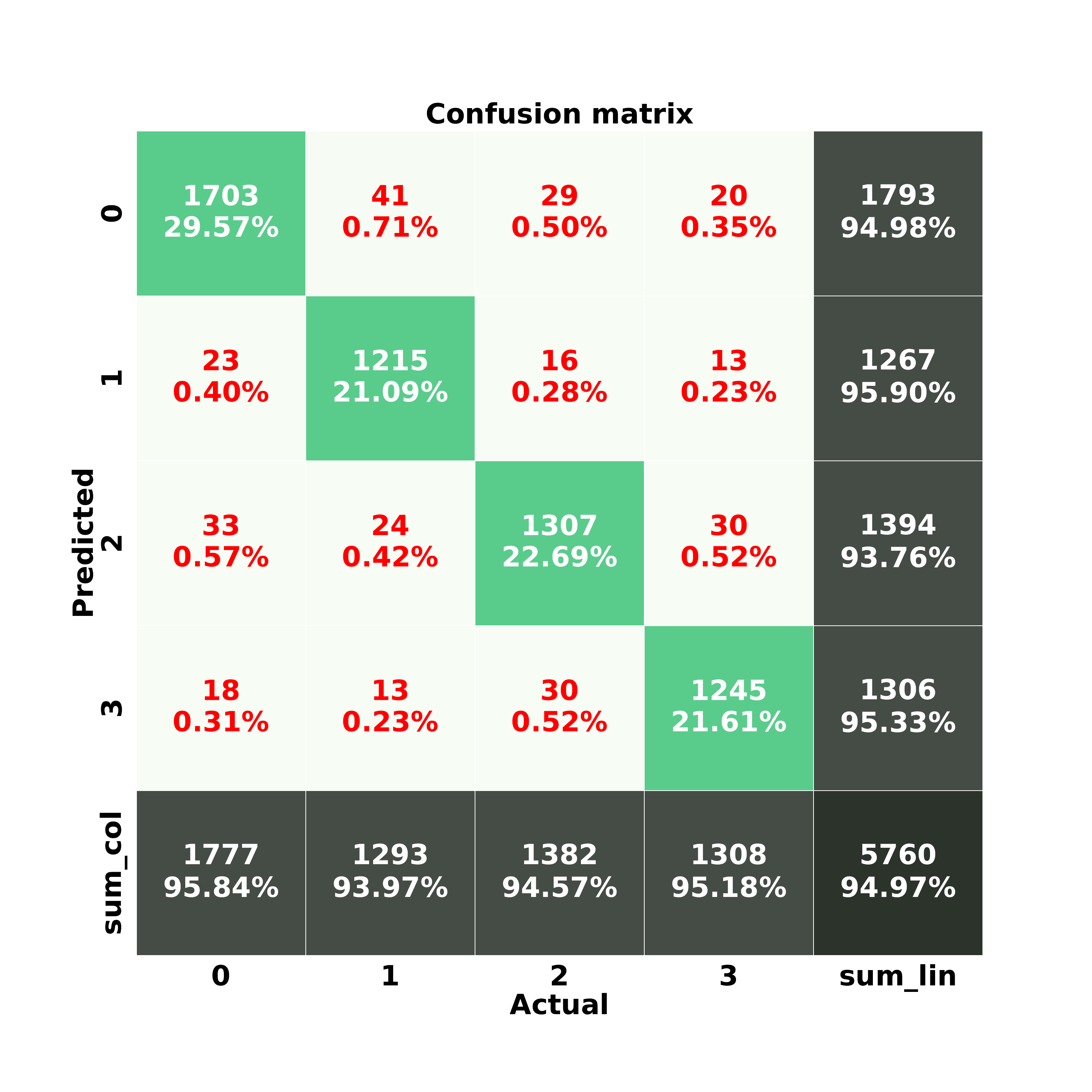}
  \caption{DEAP dataset with 4 labels (0-HVHA, 1-HVLA, 2-LVHA  and 3-LVLA)}
  \label{fig:condusiondeapfour}
\end{subfigure}

\begin{subfigure}{0.49\textwidth}
\centering
\includegraphics[width=\textwidth]{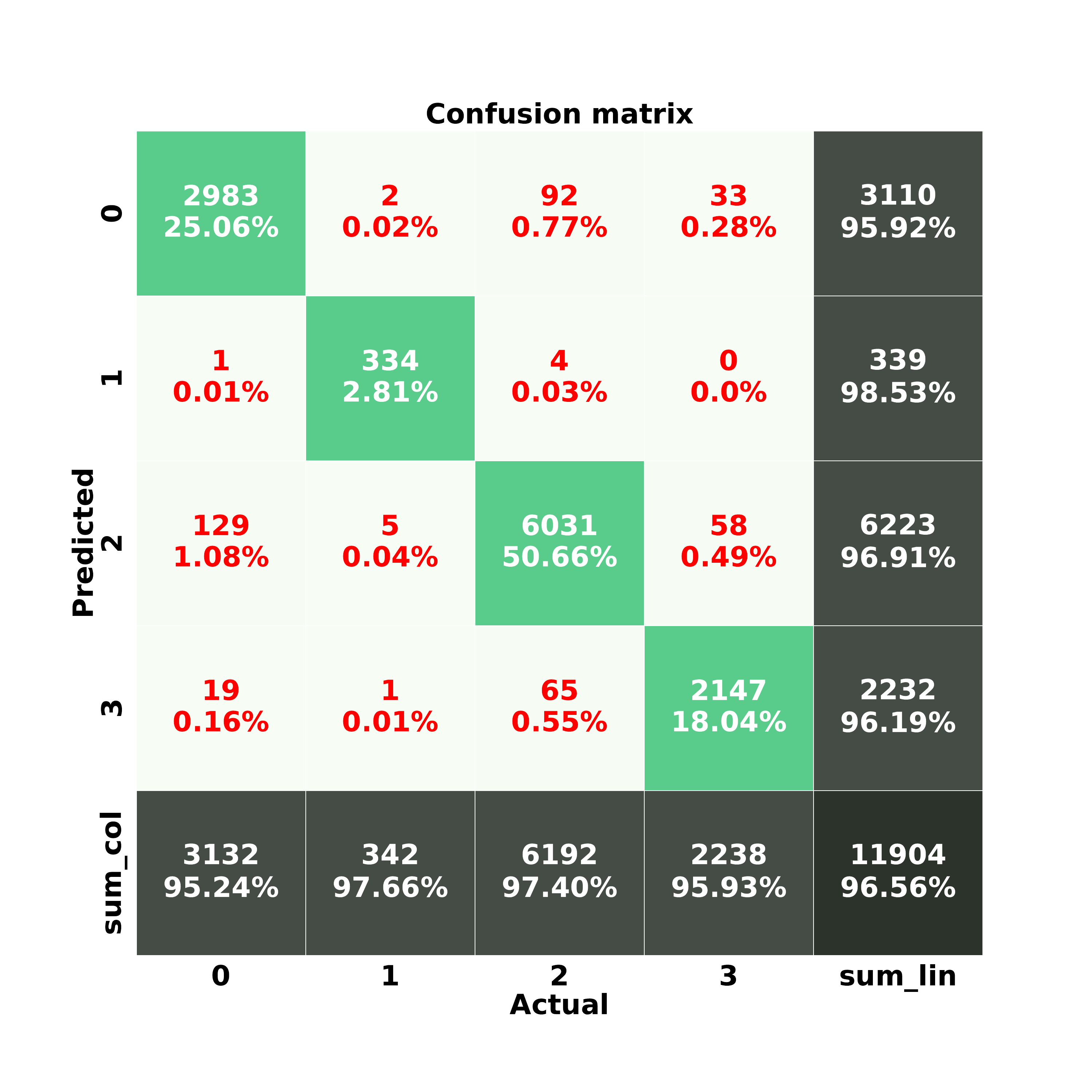}
    \caption{DENS dataset with 4 labels (0-HVHA, 1-HVLA, 2-LVHA  and 3-LVLA)}
    \label{fig:condusiondensfour}
\end{subfigure}


\caption{Comparison of Confusion matrices for DEAP and DENS datasets over Valence-Arousal space. This space is divided into four classes and assigned a label to it (0-HVHA, 1-HVLA, 2-LVHA  and 3-LVLA). Abbreviations of the terms- V:Valence; A:Arousal; L:Low; H:High.}
\label{fig:confDeapSeed}
\end{figure}

\begin{figure}
\centering


\begin{subfigure}{0.49\textwidth}
\centering
  \includegraphics[width=\textwidth]{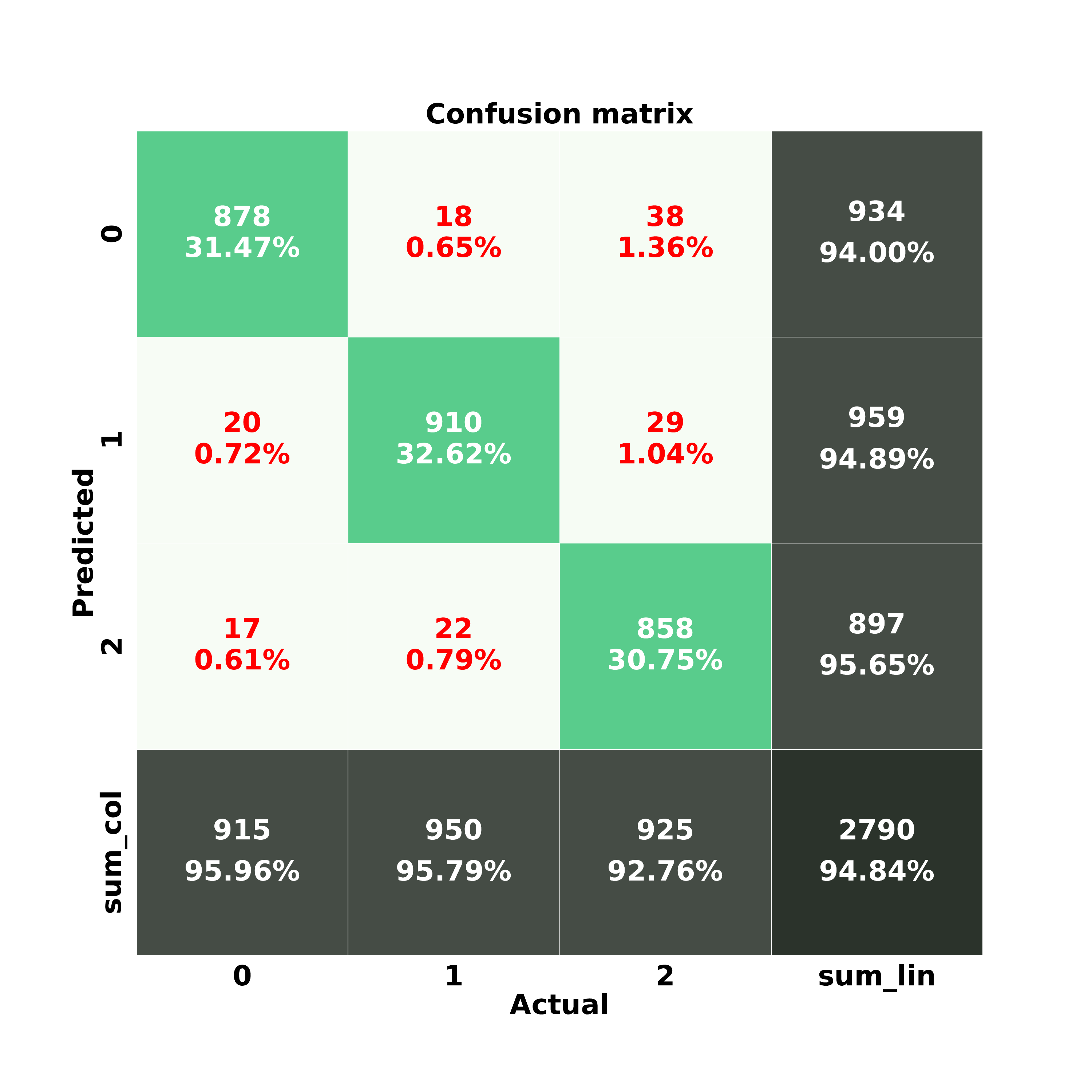}
  \caption{SEED dataset with 3 labels (0 for neutral, 1 for positive and 2 for negative)}
  \label{fig:condusionseedthree}
\end{subfigure}

\begin{subfigure}{0.49\textwidth}
\centering
\includegraphics[width=\textwidth]{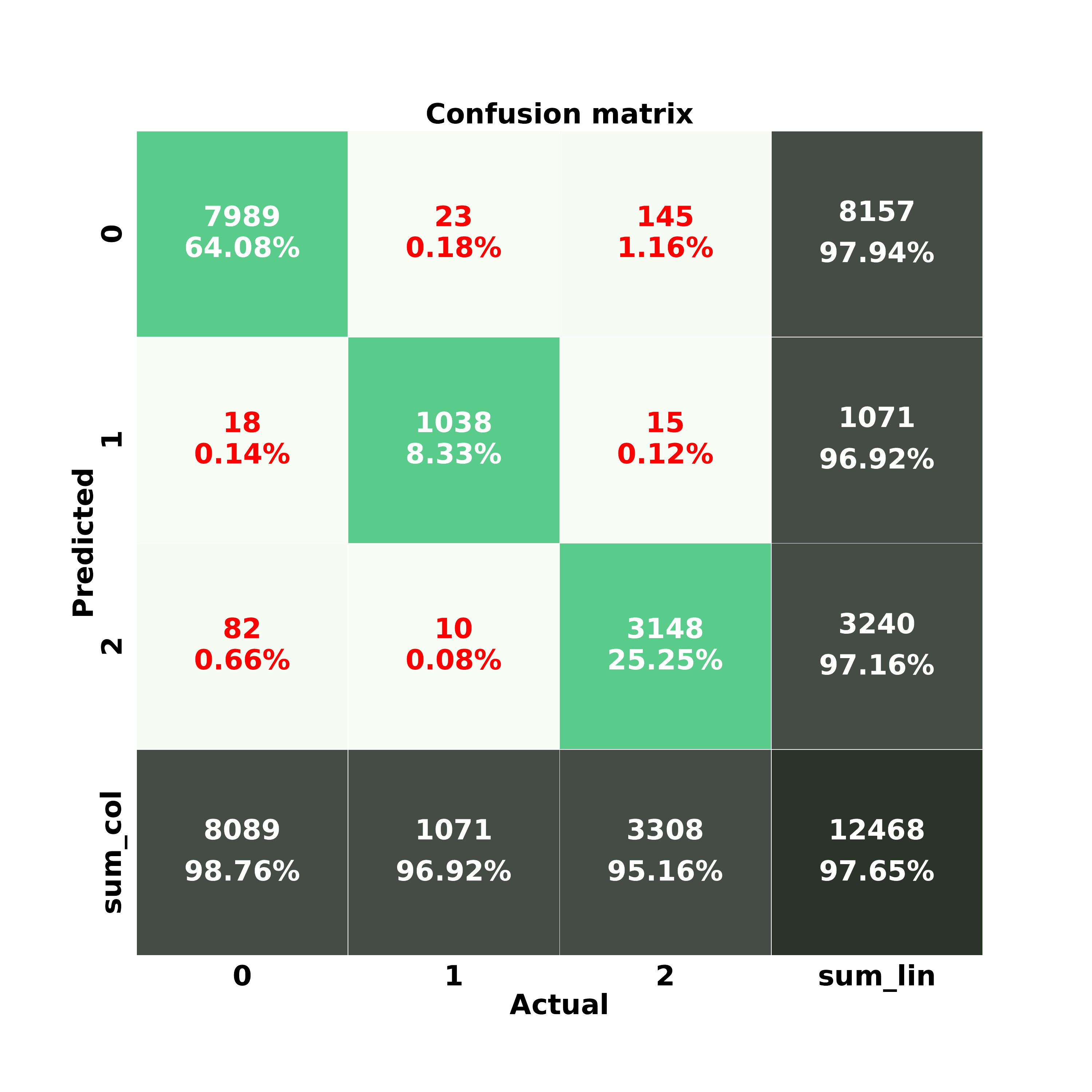}
  \caption{DENS dataset with 3 labels namely, 0 for low-valence, 1 for non-emotional data and 2 for high-valence.}
  \label{fig:condusiondensthree}
\end{subfigure}
\caption{Comparison of Confusion matrices for SEED and DENS datasets over Valence space. This space is divided into three classes (SEED dataset provided data with three classes, while DENS data is divided into three classes based on the valence ratings provided by the participants)and assigned a label to it as follows:
For SEED:0 for neutral, 1 for positive and 2 for negative
For DENS: 0 for low-valence (valence ratings ranges from 1-4.5), 1 for non-emotional data (valence ratings ranges from 4.5-5.5) and 2 for high-valence (valence ratings ranges from 5.5-9).}
\label{fig:ConfMatrix}
\end{figure}

The confusion matrix for DEAP, SEED and DENS datasets are shown in \ref{fig:condusiondeapfour}, \ref{fig:condusionseedthree}, \ref{fig:condusiondensfour} and \ref{fig:condusiondensthree}. In the confusion matrix shown, each cell contains data on the number of population and percentage of the population. The X-axis represents actual labels and the Y-axis represents predicted labels by the classifier. There is an additional row and a column in this confusion matrix, where the last column (sum\_lin) data mention precision for that label, and the last row data (sum\_col) mention the recall. The intersection cell of the last column and the last row shows accuracy and the total number of supports. The diagonal of the matrix represents the correctly identified label. 

\vspace{5 pt}

\subsection{Comparison between DEAP and DENS:}

\vspace{5 pt}
We have used repeated K-Fold cross-validation with K = 5 and number of repeats = 5 so generated 25 accuracies for DENS and DEAP. For label classification we have used V-A space (HVHA, HVLA, LVLA, LVHA). Comparison between DEAP and DENS is mentioned in Table \ref{tab:Deapdensmean}.

\vspace{5 pt}
\begin{table}[ht]
    \centering
    \begin{tabularx} {0.5\textwidth}{ 
  | >{\centering\arraybackslash}X 
  | >{\centering\arraybackslash}X | }
 \hline
 \textbf{Dataset} & \textbf{Mean F1 score (in \%)} \\
 \hline
 DEAP & 95.65 ($\pm$ 0.38)  \\
 \hline
 DENS & 96.82 ($\pm$ 0.18)  \\
\hline
\end{tabularx}
    \caption{DEAP vs DENS with mean F1 scores}
    \label{tab:Deapdensmean}
\end{table}

\begin{figure}
\centering
\begin{subfigure}{0.49\textwidth}
  \includegraphics[width=\linewidth, height=0.8\linewidth]{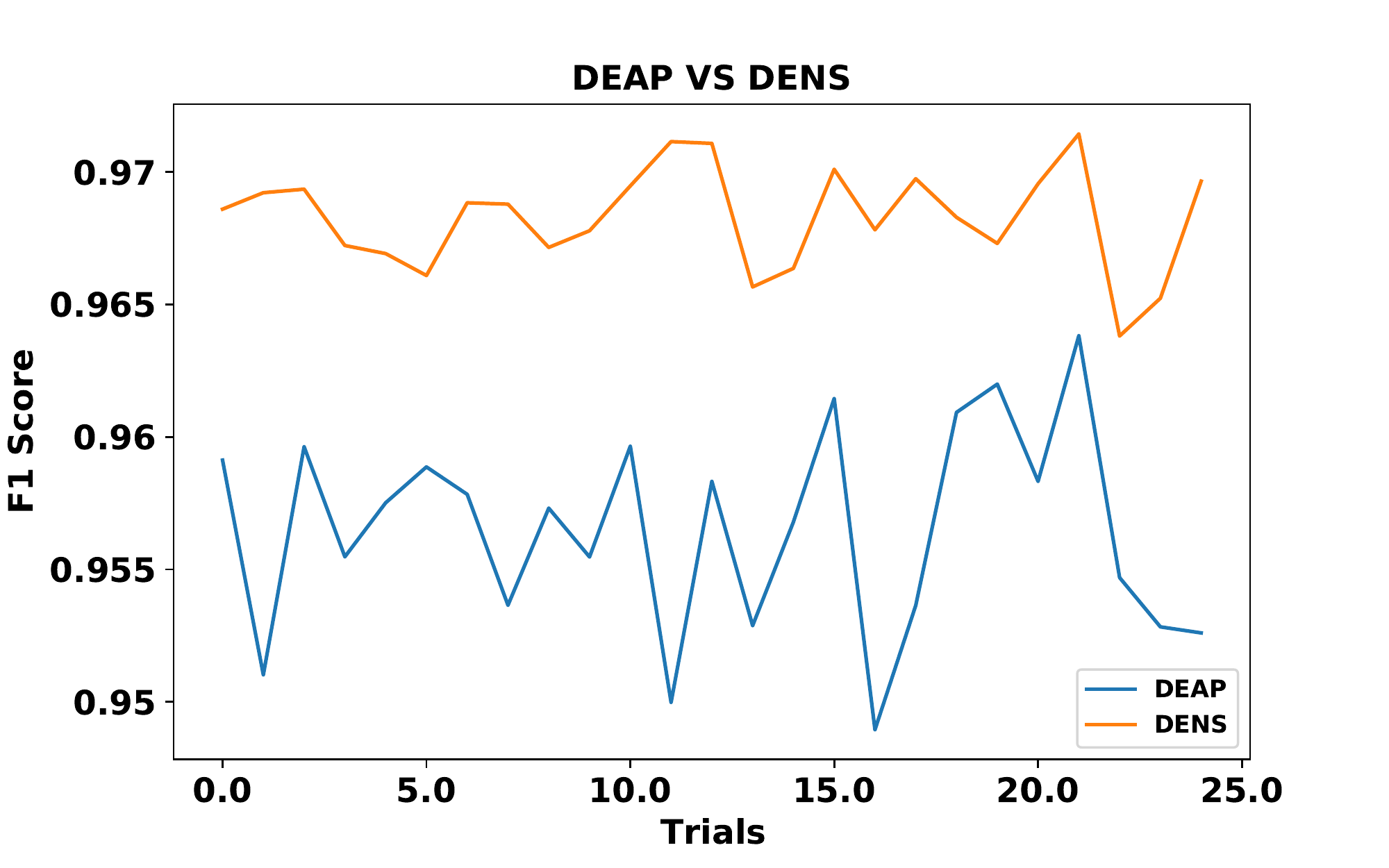}
  \caption{DEAP vs DENS}
  \label{fig:deapdenscmpr}
\end{subfigure}  
\begin{subfigure}{0.49\textwidth}
\centering
  \includegraphics[width=\linewidth, height=0.8\linewidth]{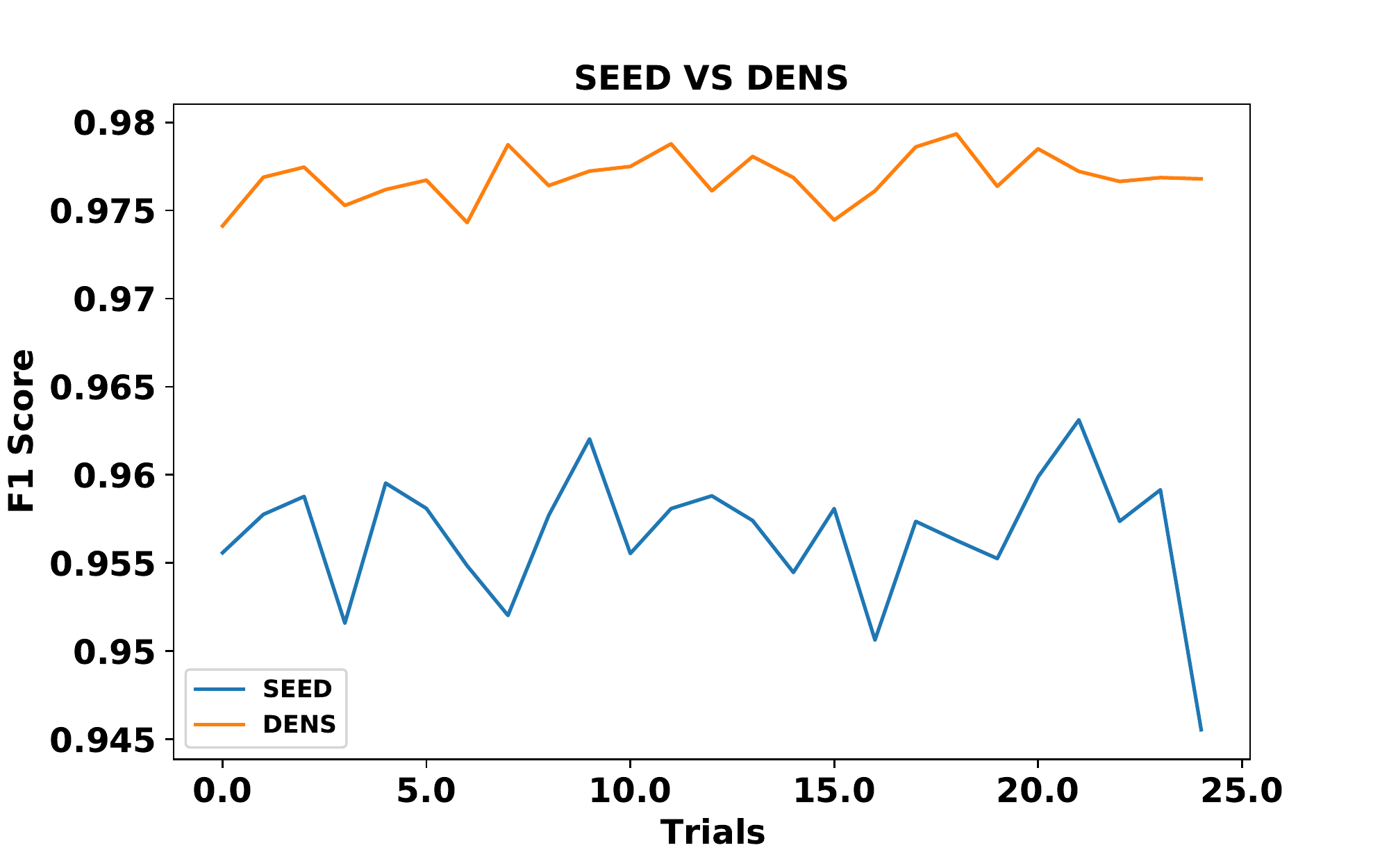}
  \caption{SEED vs DENS}
  \label{fig:seeddenscmpr}
\end{subfigure}
\caption{F1 scores of DENS comparing with DEAP and SEED datasets for all the 25 trials.}
\label{fig:accGraphs}
\end{figure}

\vspace{5 pt}

Figure \ref{fig:deapdenscmpr} shows a F1 score comparison between the DEAP and DENS dataset per trial. Using t-test statistical testing, the 25 F1 scores of DEAP dataset (M = 95.65\%, SD = 0.38\%) compared with the 25 F1 scores of DENS dataset (M = 96.82\%, SD = 0.18\%), DENS dataset shows better results with absolute t (35) = 13.54, $p<0.0001$, d estimate: -13.70 (large), 95 percent confidence interval:  [-16.51 -10.89].  
\vspace{5 pt}

\subsection{Comparison between SEED and DENS:}

\vspace{5 pt}
For SEED vs DENS comparison, label classification we have used 3 labels on valence scale. Comparison between SEED and DENS result is mentioned in Table \ref{tab:seeddensmean}.

\vspace{5 pt}
\begin{table}[ht]
    \centering
    \begin{tabularx} {0.5\textwidth}{ 
  | >{\centering\arraybackslash}X 
  | >{\centering\arraybackslash}X | }
 \hline
 \textbf{Dataset} & \textbf{Mean F1 scores (in \%)} \\
 \hline
 SEED & 95.65 ($\pm$ 0.37)  \\
 \hline
 DENS & 97.68 ($\pm$ 0.13)  \\
\hline
\end{tabularx}
    \caption{SEED vs DENS with mean F1 scores}
    \label{tab:seeddensmean}
\end{table}

\vspace{5 pt}

Figure \ref{fig:seeddenscmpr} shows a F1 score comparison between the SEED and DENS dataset per trial.
Using t-test statistical testing, the 25 F1 score of SEED dataset (M = 95.65\%, SD = 0.37\%) compared with the 25 F1 score of DENS dataset (M = 97.68\%, SD = 0.13\%), DENS dataset shows better results with absolute t (31) = 25.466, $p<0.0001$, Cohen’s d estimate: -11.37 (large), 95 percent confidence interval: [-13.73 -9.02]. 

\section{Discussion}
In this work, we captured emotional experiences within the ecologically valid naturalistic environment with a precise temporal marker than any study to date. As per the recent theories, emotional experience is a constructing phenomenon which involves networks of the brain, including the default mode network, salience network, and fronto-parietal network. These networks are not specific to emotional experiences. In fact, these networks are domain-general networks which are involved in perception (in general). Though, the connectivity among these networks might not be the same in different perceptions which are apparently shown in our previous work \cite{mishra2022dynamic}. In addition, different from normal perception, emotional experiences involve changes in body physiology\cite{mishra2022cardiac}. Putting together the above-mentioned ideas from recent results hints that the emotional experiences can be easily confused with the other perceptions, which might not be an emotional experience. 

One of the major concerns is the mind-wandering activity while using the film stimuli. In the previous research, the whole stimulus is considered to elicit a single emotional experience. And the duration of the stimulus varied from seconds to minutes. Research shows that averaging the participant's feedback for the whole duration of the stimulus might not be correctly capturing emotional experience (in particular) \cite{saarimaki2021naturalistic}. Hence, it is important to know the duration of the emotional experience without compromising the ecological validity of the stimuli. 

The main idea behind this work is that if we can capture the temporal marker of emotional experience within the real-life resembling environment, we might achieve better accuracy than the accuracy achieved to date with other datasets lacking the information about time. Although, due to the limited number of subjects, we didn't go for the subject-independent classification for now. Though, in future, we will be collecting more data to mitigate this limitation. 

In our results, we observed that the same hybrid deep learning model on our dataset outperformed other datasets, including benchmark datasets like DEAP and SEED. Classification of DEAP data into four labels, including HVHA, LVHA, LVLA, and HVLA, resulted in 94.74\% accuracy. At the same time, the classification of DENS data into four labels resulted in 96.38\% accuracy. The significance testing showed that even with the multiple iterations, the classification accuracy was significantly higher for our data. 






To the date, most of the work on emotion recognition applied different shallow machine learning and deep learning techniques using many different configurations of input data including, spectrogram, raw signals, statistical features, variational mode decomposition (VMD), empirical mode decomposition (EMD), functional connectivity based features, fractal features and so on. However, still, the recognition of emotion from EEG stands as a problem. Most of the works on emotion recognition have used some benchmark datasets, including DEAP, SEED, AMIGO, MAHNOB-HCI and so on. Though, most of the emotion classification works revolves around DEAP and SEED datasets \cite{koelstra2011deap}. 


In \cite{lee2014classifying}, emotional states are classified by means of EEG-based functional connectivity patterns. Forty participants viewed audio-visual film clips to evoke neutral, positive (one amusing and one surprising) or negative (one fear and one disgust) emotions. Correlation, coherence, and phase synchronization are used for estimating the connectivity indices. They stated significant differences among emotional states. A maximum classification rate of 82\% was reported when the phase synchronization index was used for connectivity measure.


The classes considered in the study are elementary. We suspect that with the increasing number of emotional classes, which includes not only basic classes but complex emotions as well, taking the long duration signal without a temporal marker may not be able to categorize emotional classes. The reason is that there are fewer chances for a movie stimulus to have positive as well as a negative emotional experience in the same stimuli, but it is certainly possible that it can have more than one positive or more than one negative feeling in the movie.


\section{Conclusion}
The work presented in this article is based on the concept that emotion is a short-lived phenomenon which might last for very few seconds. Hence, using long-duration EEG signals recorded during emotional stimulus watching might not contain emotional information for the whole duration. Therefore, we hypothesized that using only the duration of the signal where an emotional event is reported without compromising the ecological validity of the stimuli will contain more emotional information. To test the hypothesis, we designed an EEG experiment which uniquely marks the duration of the emotional event in the continuous recording of brain waves using EEG. We performed deep learning analysis using a hybrid CNN and LSTM model and found results that significantly favoured our hypothesis. In this work, we saw the problem with a different aspect which have not attracted the attention of the researcher. We suggest that future research on emotion recognition should adapt our approach to collect more such kinds of data so that emotion recognition using EEG can go beyond the recognition of basic emotions using facial features or body physiology toward more complex emotions.

\bibliographystyle{unsrt}  
\bibliography{references}

\end{document}